\documentclass{article}

\usepackage{amsmath}
\usepackage{graphics,graphicx}
\usepackage{authblk}
\usepackage{amsfonts}

\usepackage[round,authoryear]{natbib} 
\usepackage{algpseudocode,algorithm}
\usepackage{amsmath,amssymb, amsbsy}
\usepackage{epsfig,graphics,graphicx}
\usepackage{url}

\newtheorem{thm}{Theorem}[section]

\newtheorem{lem}{Lemma}[section]

	\usepackage{lineno}
	
\def\mR{\mathbb{R}}
\def\tr{\mbox{tr}}

\def\wvec{\mbox{vec}}

\newcommand{\trans}{^{\mbox{\tiny{T}}}}
\DeclareMathOperator*{\argmin}{arg\,min} 
\begin{document}
	\title{A Fast Iterative Algorithm for High-dimensional Differential Network }
	\author[1,2]{Zhou Tang  \thanks{johntan@sjtu.edu.cn}}
	\author[1,2]{Zhangsheng Yu\thanks{Co-first author: yuzhangsheng@sjtu.edu.cn}}
	\author[2]{Cheng Wang  \thanks{Corresponding Author: chengwang@sjtu.edu.cn}}
	\affil[1]{Department of Bioinformatics and Biostatistics, Shanghai Jiao Tong University, Shanghai, 20040,  China.}
	\affil[2]{Department of Mathematics, Shanghai Jiao Tong University, Shanghai, 20040,  China.}
	\date{\today}
	\maketitle
	\begin{abstract}
	Differential network is an important tool to capture the changes of conditional correlations under two sample cases.  In this paper, we introduce a fast iterative algorithm to recover the differential network for high-dimensional data. The  computation complexity of our algorithm is linear in the sample size and the number of parameters, which is optimal in the sense that it is of the same order as  computing two sample covariance matrices.  The proposed method is appealing for high-dimensional data with a small sample size. The experiments on simulated and real data sets show that the proposed algorithm outperforms other existing methods.
		
	\end{abstract}

\section{Introduction}

Covariance matrices which describe the correlations between covariates play an important role in multivariate statistical analysis. For high-dimensional data where the number of covariates is large, it is challenging to estimate the covariance matrix.  In literature, a large number of statistical methods have been proposed to estimate the covariance matrix \citep{bickel2008regularized, rothman2009generalized, cai2011adaptive} or its inverse which is usually called as the precision matrix \citep{meinshausen2006high, Friedman2007Sparse, cai2011constrained, zhang2014Sparse}.  More details can be found in  recent review works by \cite{tong2014estimation} or \cite{fan2016overview}.

In this work, we study the covariance structure for the two-sample cases. Suppose that we have observations from two groups of subjects: $X_1,\dots,X_{n_1}$ and $Y_1,\dots,Y_{n_2}$ whose population covariance matrices are $\Sigma_1$ and $\Sigma_2$, respectively. Our interest is to estimate the differential network $\Delta^\ast = \Sigma_2^{-1} - \Sigma_1^{-1}$, which is the difference between two precision matrices. In biostatistics, the differential network describes the changes of conditional interdependencies between components under different environmental or genetic conditions. 
See \cite{barabasi2004network}, \cite{bandyopadhyay2010rewiring}, \cite{barabasi2011network}, \cite{gambardella2013differential},  and \cite{zhao2014direct} for example and the references therein. Another application of the differential network is the quadratic discriminant analysis in multivariate statistical analysis \citep{anderson2003introduction}. Under the Gaussian distribution assumption, the differential network is exactly the coefficients for the interaction terms between covariates. For quadratic discriminant analysis,  it is necessary to recover the differential network \citep{li2015sparse, jiang2015quda}.

In the past decades, a large number of statistical methods have been proposed to estimate $\Delta^\ast$ which can be classified into two categories. The first one is to estimate precision matrices $\Sigma_1^{-1}$ and $\Sigma_2^{-1}$ separately, and then taking the difference yields the final estimation for the differential network. The methods for estimating a single precision matrix  \citep{meinshausen2006high, Friedman2007Sparse, cai2011constrained, zhang2014Sparse} can be used directly. The second approach is to jointly estimate precision matrices $\Sigma_1^{-1}$ and $\Sigma_2^{-1}$ \citep{Julien2011Inferring, Jian2011Joint, zhu2018multiple}. A joint loss function for the precision matrices is conducted and we can estimate the precision matrices simultaneously by penalizing the joint loss function. These methods assume that each precision matrix is sparse and can be recovered consistently which is too strong for many applications. Moreover, since our interest is only the differential network, it is not necessarily to recover each network for all the subjects.

Recently, \cite{zhao2014direct} developed a direct estimator for the differential network $\Delta^\ast$ under high-dimensional setting. Motivated by \cite{cai2011constrained}, they proposed a Dantzig-typed estimator for the high-dimensional differential network. By studying high-dimensional quadratic discriminant analysis,   \cite{jiang2015quda} proposed a LASSO-typed estimator which regularized a convex loss function with a $\ell_1$ penalized term.  Usually, the estimators of \cite{zhao2014direct} and \cite{jiang2015quda} are not symmetric and a further symmetrical step is needed for the final estimation. \cite{Huili2017Differential} conducted an one step symmetric estimator. Under mild conditions, these estimators are all shown to be consistent  by assuming that the differential network matrix is sparse. Computationally, they all used the alternating direction method of multipliers (ADMM) \citep{boyd2011distributed} to solve the optimization problems. In details, \cite{zhao2014direct} used a proximal linearizion procedure to solve the Dantzig-typed optimization problem.  The $\ell_1$ penalized problem of \cite{jiang2015quda} can be solved by standard ADMM and \cite{Huili2017Differential} proposed a two step ADMM algorithm.

For high-dimensional data where $p \gg n$, the computational complexity of \cite{zhao2014direct} is $O(p^4)$ while \cite{jiang2015quda} and \cite{Huili2017Differential} improved the complexity to $O(p^3)$. 
In this paper, we introduce a fast iterative shrinkage-thresholding algorithm \citep{beck2009fast} to minimize loss functions defined in \cite{Huili2017Differential} and \cite{jiang2015quda}. The computational complexity of the new method is improved to around $O(n p^2)$, which is the same as computing the two sample covariance matrices. Moreover, the proposed iterative shrinkage-thresholding algorithm is a first order method which is based on the gradients and avoids calculating the inverse of matrices.  The theoretical convergence rate is also given in this paper. Lastly, simulation studies and real data analysis justify the advantages of our algorithm. An R package of our method has been developed and is available at \url{http://math.sjtu.edu.cn/faculty/chengwang/pub.html}.

The rest of the paper is organized as follows. In Section 2, we introduce the loss functions in existing methods and propose the new algorithm. Evaluations in simulated data are presented in Section 3 and in Section 4, the algorithm is applied to two real data sets to demonstrate its performance. The theoretical convergence rate of the algorithm are proved in Appendix.

\section{Main Results}
For any real matrix $A$, we shall use $\|A\|_2=\sqrt{\tr(AA\trans)}$ to denote its Frobenius norm, $\|A\|$ to denote its spectral norm and $\|A\|_1$ denotes the sum of the absolute values of $A$. 
\subsection{Existing Methods}
Our interest is to estimate the differential network $\Delta^\ast=\Sigma_2^{-1}-\Sigma_1^{-1}$ which is defined as the difference between two precision matrices.  Noting
\begin{align}
\Delta^\ast=\Sigma_2^{-1}-\Sigma_1^{-1}=\Sigma_1^{-1}(\Sigma_1-\Sigma_2)\Sigma_2^{-1},
\end{align}
we can get 
\begin{align*}
\wvec{(\Delta^\ast)}=(\Sigma_2^{-1}\otimes \Sigma_1^{-1}) \wvec{(\Sigma_1-\Sigma_2)}=(\Sigma_2 \otimes \Sigma_1)^{-1} \wvec{(\Sigma_1-\Sigma_2)},
\end{align*}
where $\otimes$ denotes the Kronecker product and $\wvec(\cdot)$ is the vectorization of a matrix. To estimate $\wvec{(\Delta)}$, following LASSO \citep{Tibshirani1996Regression}, we can consider the $\ell_1$ penalized estimation 
\begin{align*} 
\argmin \frac{1}{2} \beta \trans (S_2 \otimes S_1) \beta- \beta \trans \wvec{(S_1-S_2)}+\lambda \|\beta\|_1,
\end{align*}
where $S_1,S_2$ are the sample covariance matrices and $\lambda > 0$ is a tuning parameter.  Letting $\beta=\wvec{(\Delta)}$, we can get the estimation in matrix form
\begin{align}  \label{delta1}
\hat{\Delta}_1=\argmin_{\Delta \in \mR^{p \times p}} \frac{1}{2} \tr\{\Delta \trans S_1 \Delta S_2\}- \tr\{\Delta (S_1-S_2)\}+\lambda \|\Delta\|_1,
\end{align}
which is exactly the estimator proposed by \cite{jiang2015quda}. Here, the loss function 
\begin{align}\label{dloss}
L_1(\Delta)= \frac{1}{2} \tr\{\Delta \trans S_1 \Delta S_2\}- \tr\{\Delta (S_1-S_2)\},
\end{align}
is convex with respect to $\Delta$ which is appealing for optimization. Generally, the estimation $\hat{\Delta}_1$ is not symmetric and a further symmetrization is needed to obtain the final estimator.  \cite{Huili2017Differential} considered a symmetric loss function
\begin{align}\label{sloss}
L_2(\Delta)=\frac{L_1(\Delta)+L_1(\Delta \trans)}{2},
\end{align} 
and proposed a symmetric estimation
\begin{align} 
\hat{\Delta}_2=\argmin_{\Delta \in \mR^{p \times p}}& \frac{1}{4} \tr\{\Delta \trans S_1 \Delta S_2\}+ \frac{1}{4} \tr\{\Delta \trans S_2 \Delta S_1\}\nonumber \\
&- \tr\{\Delta (S_1-S_2)\}+\lambda \|\Delta\|_1.  \label{delta2}
\end{align}

Theoretically, assuming $\Delta^\ast$ is sparse, \cite{jiang2015quda} and \cite{Huili2017Differential} show that $\hat{\Delta}_1$ and $\hat{\Delta}_2$ are consistent estimators for the true differential network $\Delta^\ast$. Computationally,   the loss functions $L_k(\Delta),k=1,2$ are convex functions and standard ADMM \citep{boyd2011distributed} can be used to solve the estimation \eqref{delta1} or \eqref{delta2}. In details, for the loss function $L(\Delta)=L_1(\Delta)$ or $L_2(\Delta)$, the augmented Lagrangian function is
\begin{align*}
L(\Delta,A,B)= L(\Delta)+\rho/2 \|\Delta-A+B\|_2^2+\lambda \|A\|_1,
\end{align*}
where $\rho>0$ is the step size of ADMM. The iterative scheme of ADMM is
\begin{align*}
\Delta^{k+1}&=\argmin L(\Delta,A^k,B^k)=\argmin L(\Delta)+\rho/2 \|\Delta-A^k+B^k\|_2^2,\\
A^{k+1}&=\argmin L(\Delta^{k+1},A,B^k)=\mbox{soft}(\Delta^{k+1}+B^k,\lambda/\rho),\\
B^{k+1}&=\Delta^{k+1}-A^{k+1}+B^k,
\end{align*}
where $\mbox{soft}(A,\lambda)$ is an element-wise soft thresholding operator.  The $\Delta^{k+1}$ related subproblem dominates the computation of each iteration since the other two subproblems are easy enough to calculate. Since $L(\Delta)$ is convex, it is equivalent to consider the equation
\begin{align*}
L'(\Delta)+\rho (\Delta-A^k+B^k)=0.
\end{align*}
For the estimation \eqref{delta1}, the equation is 
\begin{align} \label{eq1}
S_1 \Delta S_2-(S_1-S_2)+\rho (\Delta-A^k+B^k)=0,
\end{align} 
and solving \eqref{delta2} is related to the equation
\begin{align} \label{eq2}
& \frac{1}{2}S_1 \Delta S_2+\frac{1}{2}S_2 \Delta S_1-(S_1-S_2)+\rho (\Delta-A^k+B^k)=0.
\end{align}
The equation \eqref{eq1} can be solved efficiently with the computation complexity $O(p^3)$ and the explicit solution can be found in the Proposition 1 of \cite{jiang2015quda} or the Lemma 1 of \cite{Huili2017Differential}. For the equation \eqref{eq2}, to derive the explicit solution, it is inevitable to calculate the inverse of a $p^2 \times p^2$ matrix whose complexity is $O(p^4)$. To obtain a computationally efficient algorithm, \cite{Huili2017Differential} introduced an auxiliary iterative update  which solves the equation \eqref{eq1} twice and then combines the two solutions. In summary, the computation complexity of \cite{jiang2015quda} or \cite{Huili2017Differential} is $O(p^3)$ and an eigenvalue decomposition is necessary which will demand high computation memory.  
\subsection{New Algorithms}
In this paper, we introduce a fast iterative shrinkage-thresholding algorithm \citep{beck2009fast} to solve the penalized estimation \eqref{delta1} and \eqref{delta2}.  Compared with ADMM which needs to solve equations or equivalently calculate the inverse of matrices, the shrinkage-thresholding algorithm is a first order method which is only based on function values and gradient evaluations. 
Specially, for the estimation  \eqref{delta1} or \eqref{delta2}, the gradient can be solved efficiently and then the computational complexity can be improved to $O(n p^2 )$ where $n=n_1+n_2$. Under the high dimension small sample size setting where $p \gg n$, the  computation complexity is linear in the sample size and the number of parameters, which is the same as computing the two sample covariance matrices. 

For the optimization problem, 
\begin{align*}
\argmin_{\Delta \in \mR^{p \times p}} L(\Delta)+\lambda \|\Delta\|_1,
\end{align*}
we consider the quadratic approximation at a given point $\Delta' \in \mR^{p \times p}$, 
\begin{align}
\label{quadratic approximation}
Q(\Delta, \Delta')=  L(\Delta') +  (\Delta - \Delta') \trans  \nabla L(\Delta')+ \frac{L}{2}\| \Delta - \Delta'\|_2^2 + \lambda \| \Delta\| _1,
\end{align}
where $L> 0$ is the Lipschitz constant for the gradient  $\nabla L(\Delta)$. Since (\ref{quadratic approximation}) is a strongly convex function with respect to $\Delta$, the unique minimizer of $Q(\Delta, \Delta')$   for given $\Delta'$ is
\begin{align*}
\argmin_{\Delta \in \mR^{p \times p}} Q(\Delta, \Delta')=\mbox{soft}(\Delta'-\frac{1}{L}\nabla L(\Delta'), \frac{\lambda}{L}). 
\end{align*}
Thus, we can solve the optimization problem sequentially
\begin{align*}
\Delta_k=\argmin_{\Delta} Q(\Delta, \Delta_{k-1})=\mbox{soft}(\Delta_{k-1}-\frac{1}{L}\nabla L(\Delta_{k-1}), \frac{\lambda}{L}).
\end{align*}  
By the gradient descent algorithm for the convex functions, the sequence $\{\Delta_k\}$ converges to the solution and following \cite{beck2009fast}, we can further use an accelerated scheme to speed up the convergence. Details of the algorithm is summarized in Algorithm 1.

\begin{algorithm}
	\caption{Fast iterative shrinkage-thresholding algorithm for differential network estimation}
	\label{algorithm 1}
	\begin{algorithmic}
		\State Input: Lipschitz constant $L$ of $\nabla L(\Delta)$ and initial value $\Delta_0$; 
		\State  \textbf{Step 0.} Start from  $\Delta_{-1} = \Delta_0$, $t_0 = t_1=1$
		\State  \textbf{Step 1.} Update
		\begin{align*}
		\Delta'_{k} = \Delta_{k} + \frac{t_{k-1}-1}{t_{k}}(\Delta_k-\Delta_{k-1});
		\end{align*}
		\State  \textbf{Step 2.} Update
		\begin{align*}
		\Delta_{k+1}=
		\mbox{soft}(\Delta'_{k}-\frac{1}{L}\nabla L(\Delta'_{k}), \frac{\lambda}{L});
		\end{align*}
		\State \textbf{Step 3.} Update
		\begin{align*}
		t_{k+1} = \frac{1+\sqrt{1+4t_k^2}}{2};
		\end{align*}
		\State \textbf{Step 4.} Repeat 1 through 3 until convergence.
	\end{algorithmic}  
\end{algorithm}
The main computational burden of this algorithm is Step 2 which involves the multiplication of the matrices. Specially, we need to calculate the gradients  
\begin{align*}
\nabla L_1(\Delta)=S_1 \Delta S_2 - (S_1-S_2),~\nabla L_2(\Delta)=\frac{1}{2}S_1 \Delta S_2+\frac{1}{2}S_2 \Delta S_1 - (S_1-S_2),
\end{align*}
where $S_1,S_2,\Delta$ are all $p \times p$ matrices. If we implement the algorithm naively, the computational complexity will be $O(p^3)$  which is the same as the one of \cite{jiang2015quda} or \cite{Huili2017Differential}.  For the high-dimensional data where $p \gg n$, we have the formulas
\begin{align*}
S_1=\frac{1}{n_1} X \trans X, S_2=\frac{1}{n_2} Y \trans Y,
\end{align*}
where $X,Y$ are the $p \times n_1, p \times n_2$ centered data matrix for the two subjects, respectively. 
Then the gradient  $\nabla L(\Delta)$ can be calculated efficiently by using the facts
\begin{align*}
S_1 \Delta S_2 = \frac{1}{n_1n_2} X \trans(X\Delta Y \trans) Y,~S_2 \Delta S_1 = \frac{1}{n_1n_2} Y \trans(Y\Delta X \trans) X,
\end{align*}
where the computational complexity can be reduced to $O(n p^2)$. 

For the fast iterative shrinkage-thresholding algorithm with accelerated scheme, the sequence of function values $F(\Delta_k) \equiv L(\Delta_k) + \lambda \| \Delta_k\|_1$ can converge to the optimal value $\inf F(\Delta) $ at a linear convergence rate. That is $F(\Delta_k) - \inf F(\Delta) \leq O(1/k^2)$, which is the best iteration complexity when only first order information is used \citep{Nesterov1983a}. The following theorem gives the $O(\sqrt{L/\epsilon})$ iteration complexity for the Algorithm \ref{algorithm 1} whose proof is postponed to Appendix for the sake of clarity.
\begin{thm}
	Let $\{\Delta_k\}$ be generated by Algorithm \ref{algorithm 1} and $\Delta^* = \argmin F(\Delta)$. Then, for any $k\geq 1$,
	\begin{align*}
	F(\Delta_k) - F(\Delta^*) \leq \frac{2L\| \Delta_k-\Delta^*\| ^2}{(k+1)^2}.
	\end{align*}
	
\end{thm}

\section{Simulation Studies}
In this section, we conduct several simulations to demonstrate the performance of the proposed algorithm. In what follows, we refer to the method of \cite{zhao2014direct} as ``Dantzig", the ADMM algorithm of \cite{jiang2015quda} as ``ADMM1" and the ADMM algorithm of \cite{Huili2017Differential} as ``ADMM2". The new proposed  iterative shrinkage-thresholding algorithm are denoted as ``New1" and ``New2". All the algorithms are terminated under the same stop condition  $|F(\Delta_{k}) - F(\Delta_{k+1})| < 10^{-5} (|F(\Delta_{k})|+1)$.

For all of our simulations, we set the sample size $n_1=n_2=200$ and generate the data $X_1, \cdots, X_{n_1}$ and $Y_1, \cdots, Y_{n_2}$ from $N(0, \Sigma_1)$ and $N(0, \Sigma_2)$, respectively. The true differential network is 
\begin{align*}
\Delta^\ast=\Sigma_2^{-1}-\Sigma_1^{-1}=\begin{pmatrix}
0&-1&0&\cdots&0\\
-1&2&0&\cdots&0\\
0&0&0&\cdots&0\\
\vdots&\vdots&\vdots&\vdots&\vdots\\
0&0&0&\cdots&0\\
\end{pmatrix},
\end{align*}
and for the precision matrix $\Omega_1=\Sigma_1^{-1}$, we consider two covariance structures:
\begin{itemize}
	\item Sparse case: $\Omega_1=(0.5^{|i-j|})^{-1}_{p \times p}$. In details, $\{\Omega_{1}\}_{1,1}=\{\Omega_{1}\}_{p,p}=\frac{4}{3}$ and $\{\Omega_{1}\}_{i,i} = \frac{5}{3}$ for all other $i$. $ \{\Omega_{1}\}_{i,i+1}=\{\Omega_{1}\}_{i-1,i}=\frac{2}{3}$ and $\{\Omega_{1}\}_{i,j}=0$ for all other $i,j$;
	\item Asymptotic sparse case: $\Omega_{1}=(0.5^{|i-j|})_{p \times p}$.
\end{itemize}

Table \ref{tab1} summaries the computation time in seconds based on 10 replications where all methods are implemented in R with a PC with 3.40 GHz Intel Core i7-6700 CPU and 24GB memory. For all the methods, we solve a solution path corresponding to 50 lambda values ranging from $\lambda_{\text{max}}/2$ to $\lambda_{\text{max}}$ where $\lambda_{\text{max}}$ is the maximum absolute elements of the differential sample covariance matrices $S_1-S_2$ corresponding to the estimation $\hat{\Delta}=\textbf{0}$. From Table \ref{tab1} we can see that for large $p$, our proposed algorithms are much faster than the original ADMM methods whose complexity is $O(p^3)$ and also the ``Dantzig" method whose complexity is $O(p^4)$. Specially, based on ADMM, solving the symmetric estimation \eqref{delta2} is slower than calculating the estimation \eqref{delta1} since ADMM2 need to solve the equation \eqref{eq1} twice while ADMM1 only need to calculate \eqref{eq1} once. For the proposed shrinkage-thresholding algorithm, we can see that calculating the symmetric estimation uses less time which means the symmetry property help us get faster convergence rate.

\begin{table}[!htbp]
	\caption{The average computation time (standard deviation) of solving a solution path for the differential network. } 
	\label{tab1} 
	\resizebox{\textwidth}{!}{%
		\begin{tabular}{@{\extracolsep{5pt}} cccccc} 
			\\[-1.8ex]\hline 
			\hline \\[-1.8ex] 
			& p=100 & p=200 & p=400 & p=600 & p=800 \\ 
			\hline \\[-1.8ex] 
			&\multicolumn{5}{c}{Sparse case}\\	\\[-1.8ex] 
			\hline \\[-1.8ex] 
			Dantzig & 840.512(6.364) & $>1200$ & $>1200$ & $>1200$ & $>1200$ \\ 
			ADMM1 & 0.459(0.117) & 2.447(0.959) &  25.691(9.047) &  139.242(43.760) & 521.075(111.573) \\ 
			ADMM2 & 1.748(0.384) & 8.879(1.146) & 62.252(11.657) &  272.317(57.756) & 847.754(59.007) \\ 
			New1& 0.773(0.158) & 1.819(0.318) &  8.984(1.795) &  26.505(9.277) & 94.504(28.198) \\ 
			New2 & 0.702(0.104) & 1.797(0.205) &  8.263(1.284) &  24.919(4.594) & 65.408(9.630) \\ 
			\hline \\[-1.8ex] 
			&\multicolumn{5}{c}{Asymptotic sparse case} \\  \\[-1.8ex] 
			Dantzig & 890.869(8.029) & $>1200$ & $>1200$ & $>1200$ & $>1200$ \\ 
			ADMM1 & 0.613(0.108) & 4.584(2.136) &  44.633(11.330) &  238.78(56.298) &  1133.65(151.586) \\ 
			ADMM2 & 2.444(0.454) & 13.066(2.752) & 99.831(17.401) &  391.057(74.248) & 1026.663(126.072) \\ 
			New1& 0.745(0.147) & 2.234(0.322) &  11.496(2.270) &  36.629(12.586) & 150.954(17.403) \\ 
			New2 & 0.684(0.094) & 2.001(0.302) &  9.932(1.042) &  29.175(5.680) & 105.402(20.471) \\ 
			\hline \\[-1.8ex] 
	\end{tabular} }
\end{table} 

\begin{figure}[!ht]
	\centerline{
		\begin{tabular}{ccc}
			\psfig{figure=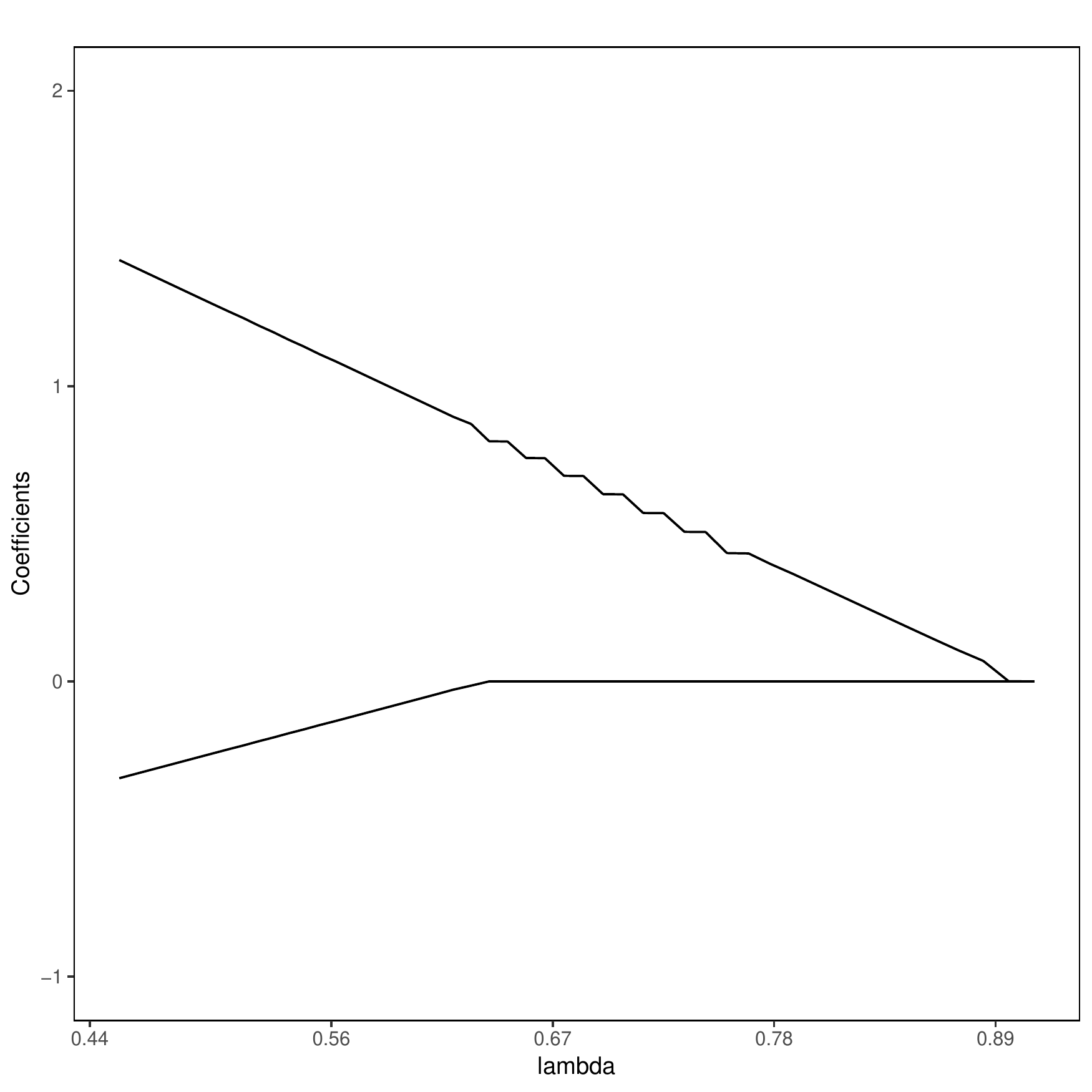,width=1.5in,height=2in,angle=0} &
			\psfig{figure=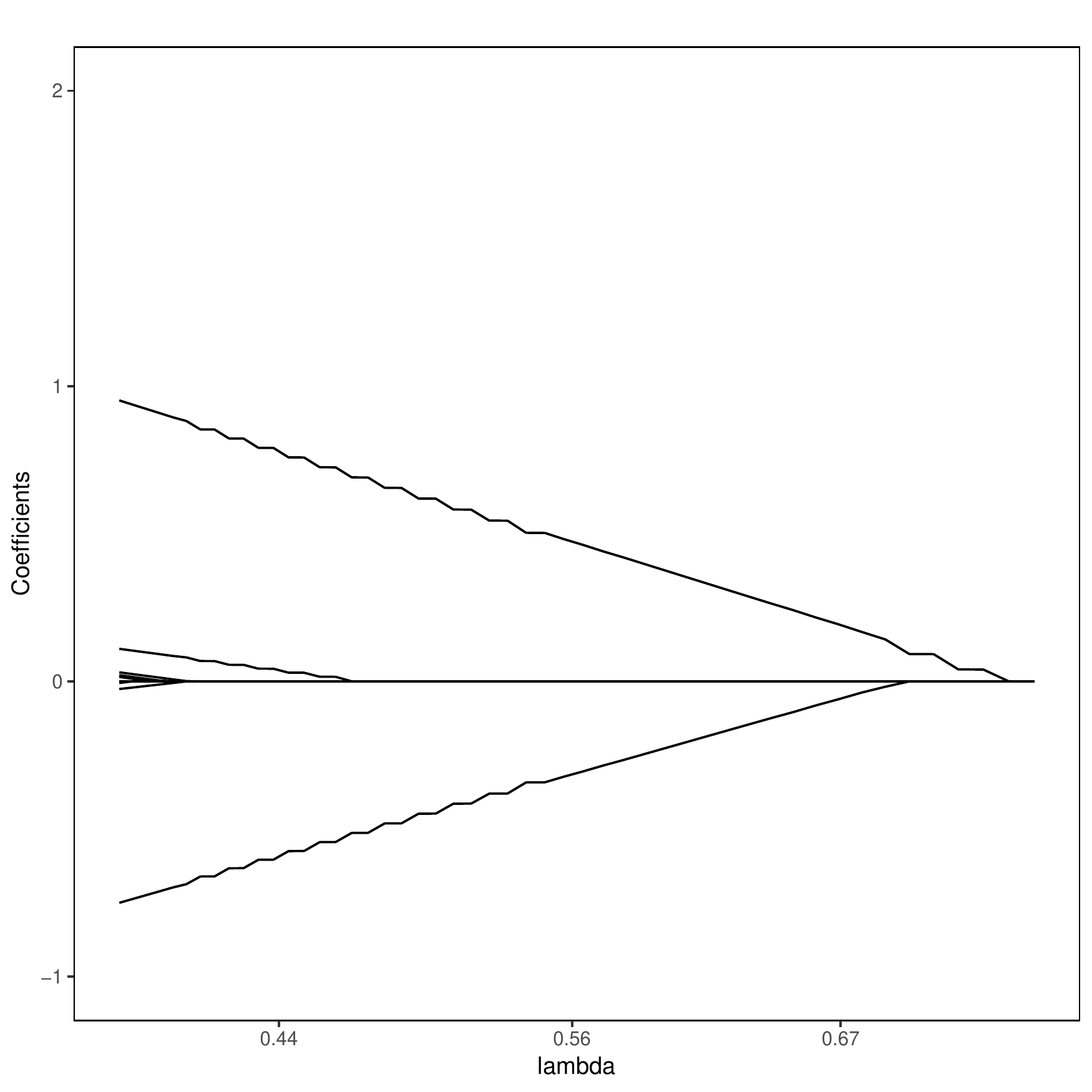,width=1.5in,height=2in,angle=0} &
			\psfig{figure=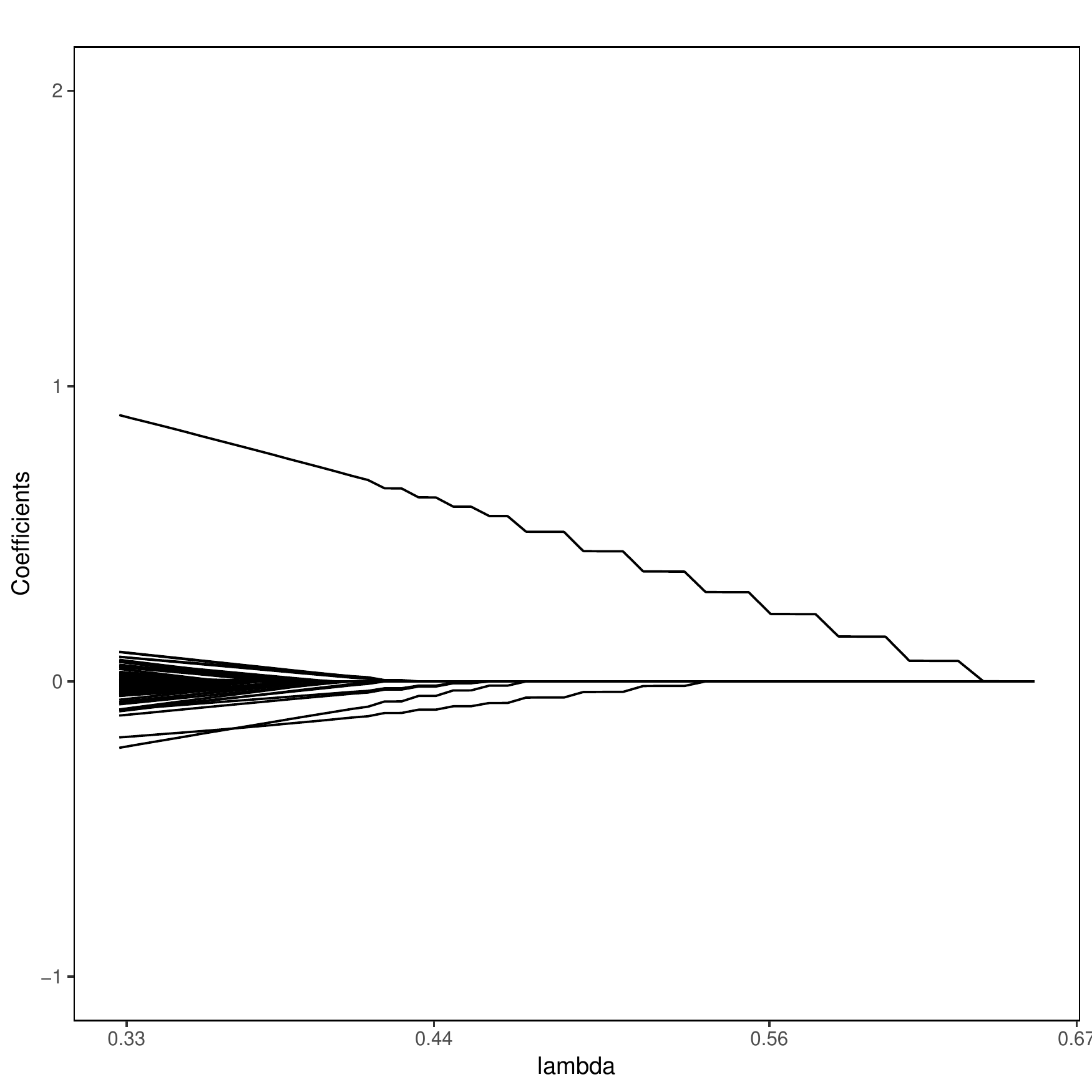,width=1.5in,height=2in,angle=0} \\
			(A): $p=100$ & (B): $p=200$ & (C): $p=400$\\
			\psfig{figure=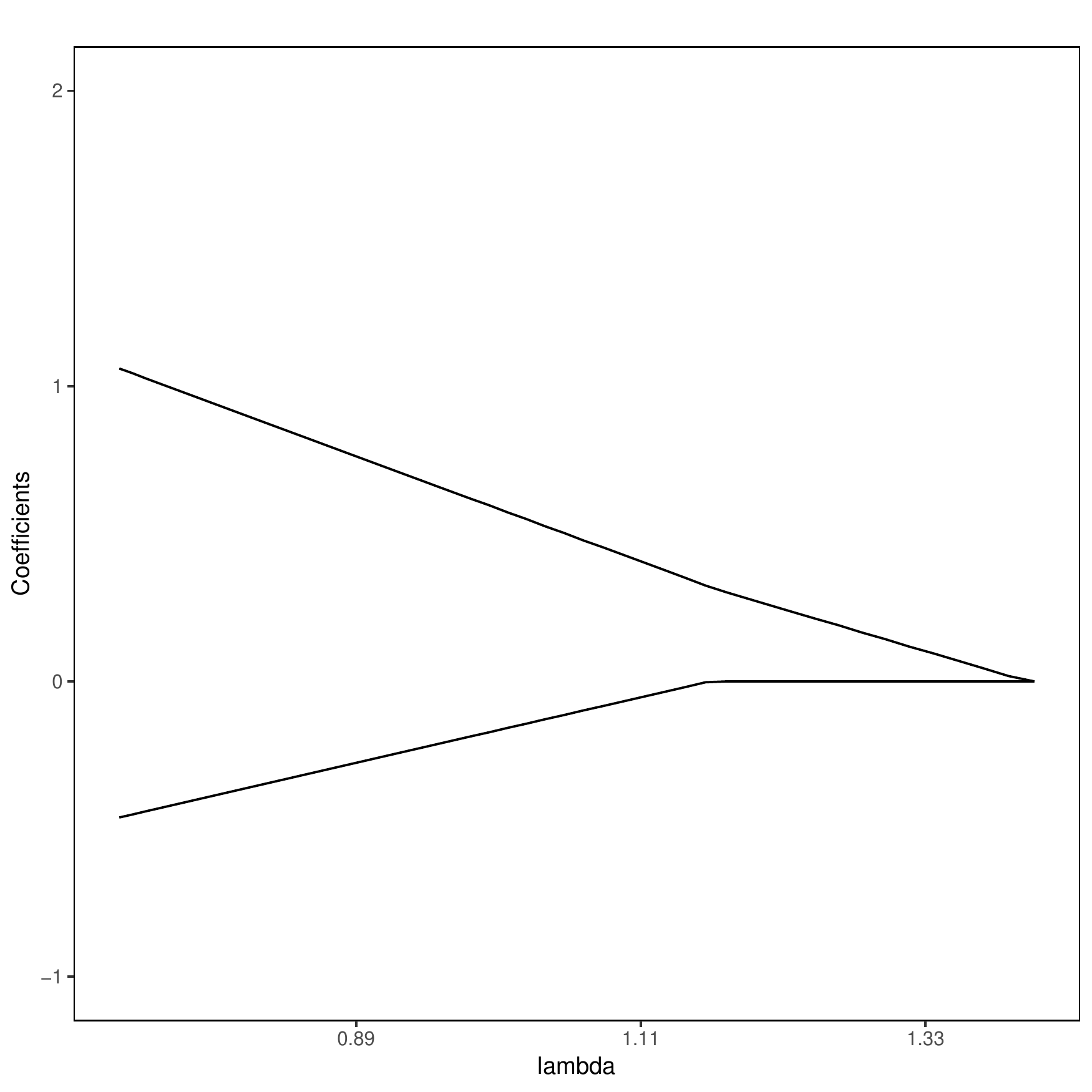,width=1.5in,height=2in,angle=0} &
			\psfig{figure=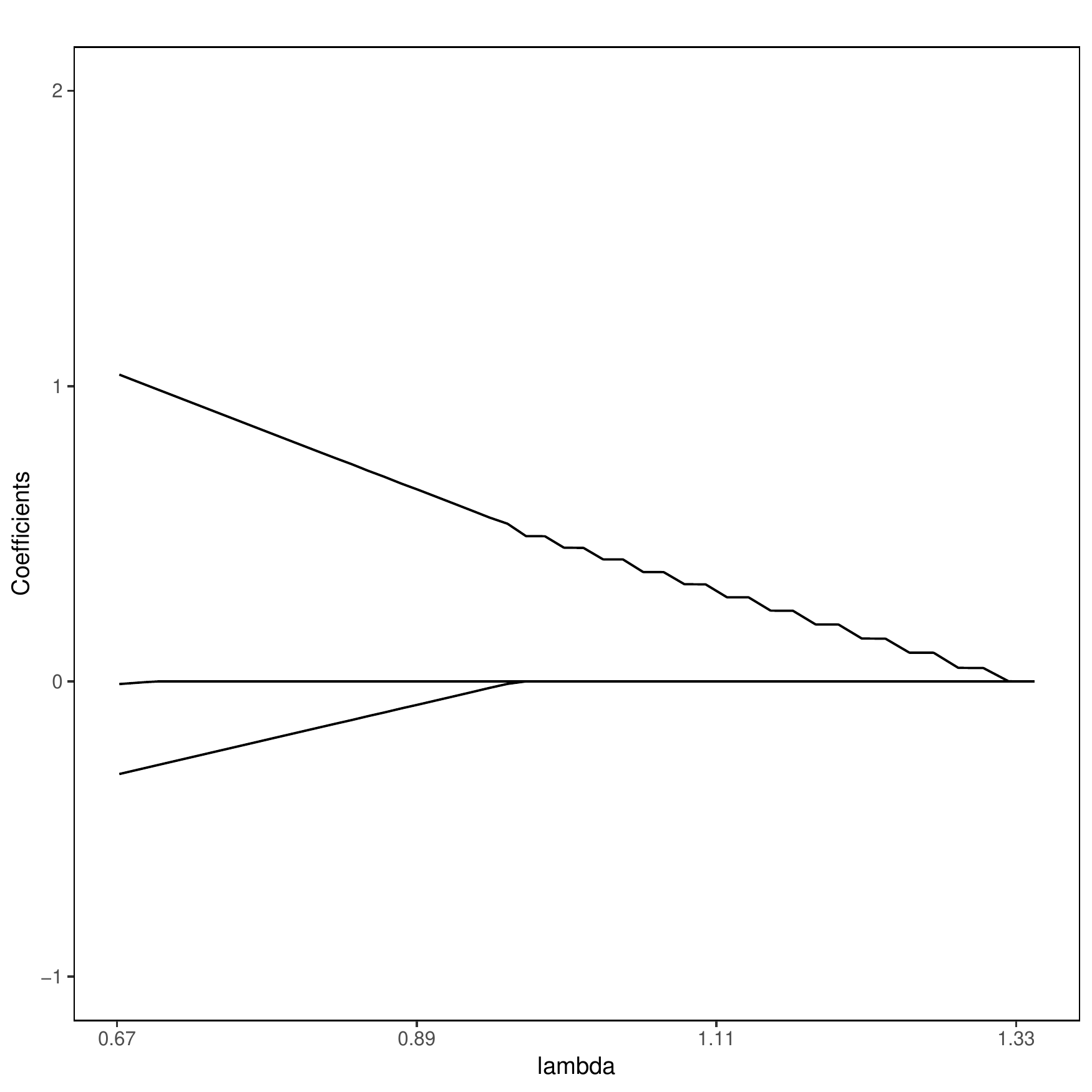,width=1.5in,height=2in,angle=0} &
			\psfig{figure=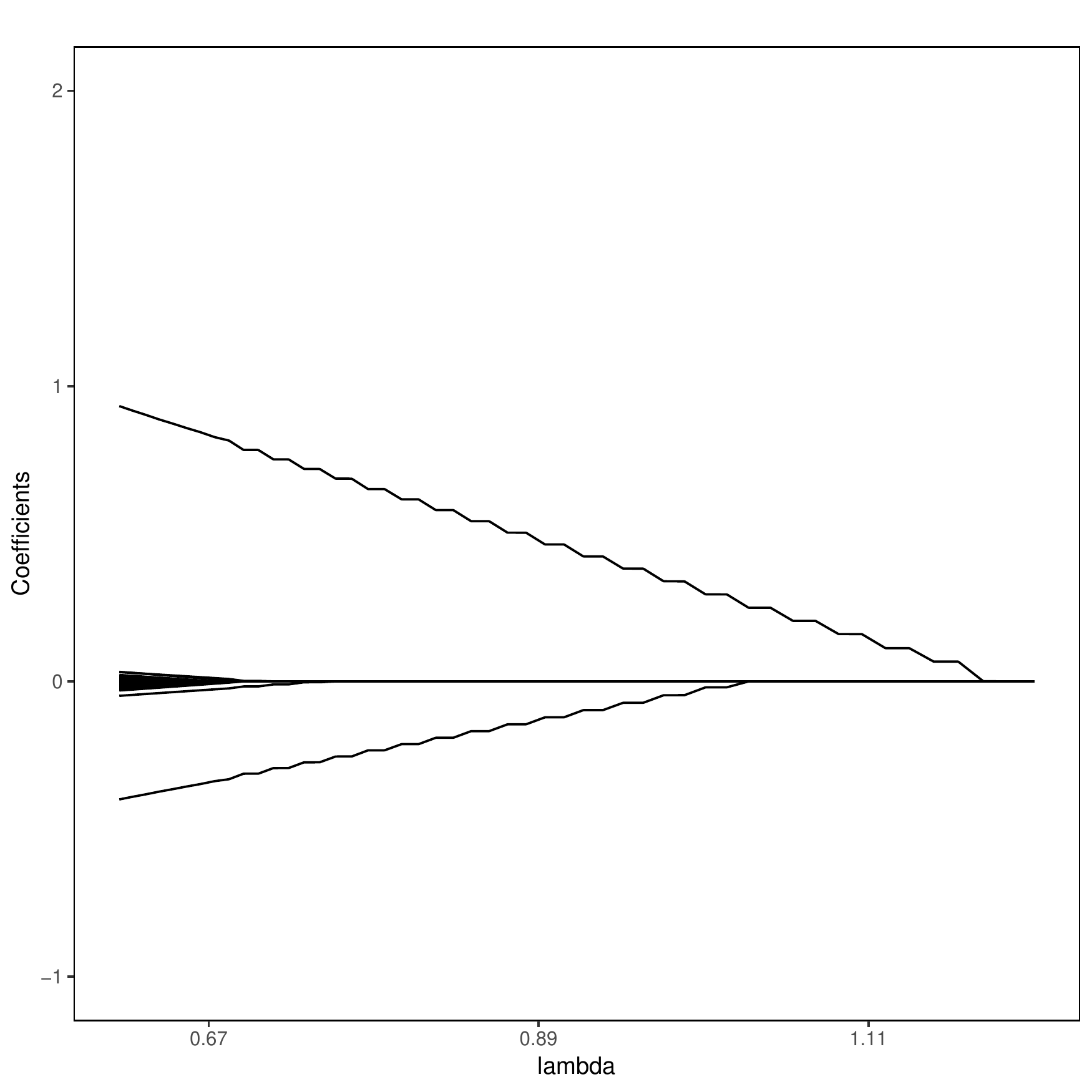,width=1.5in,height=2in,angle=0} \\
			(D): $p=100$ & (E): $p=200$ & (F): $p=400$\\
		\end{tabular}
	}
	\caption{The solution paths for different data dimension $p$ where the top panels are results for sparse cases and the bottom panels are results for asymptotic sparse cases.}
	\label{fig1}
\end{figure}

Figure \ref{fig1} shows the solution paths of the symmetric estimation for the sparse case and the asymptotic sparse case with different data dimension $p$. We can see that the $\ell_1$ penalized methods \eqref{delta2} does can recover the support of the differential network  when the tuning parameter is suitably chosen. 

\section{Real Data Analysis}
In this section we apply our algorithm to two real data sets.
\subsection{Spambase Data Set}
In this example, we model the differential network of spam and non-spam emails. The data is publicly  available at \url{https://archive.ics.uci.edu/ml/datasets/spambase}, which includes 1813 spam emails and 2788 non-spam emails. The data set collects 56 attributes including the frequency of the words and the characters and also the length of the uninterrupted sequences of capital letters. More details can be found in the website.
\begin{figure}[!ht]
	\centerline{
		\begin{tabular}{cc}
			\psfig{figure=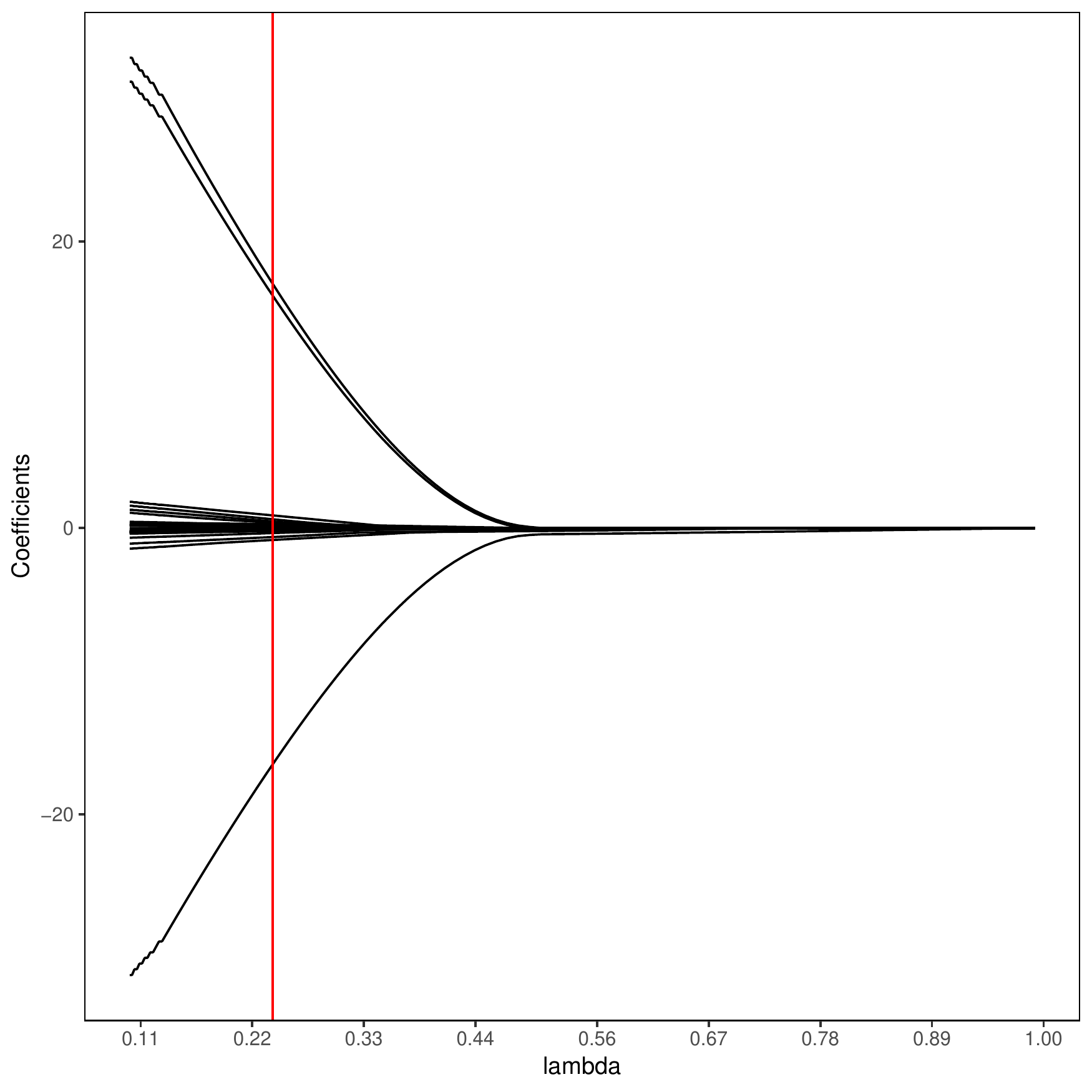,width=2.25in,height=2.5in,angle=0} &
			\psfig{figure=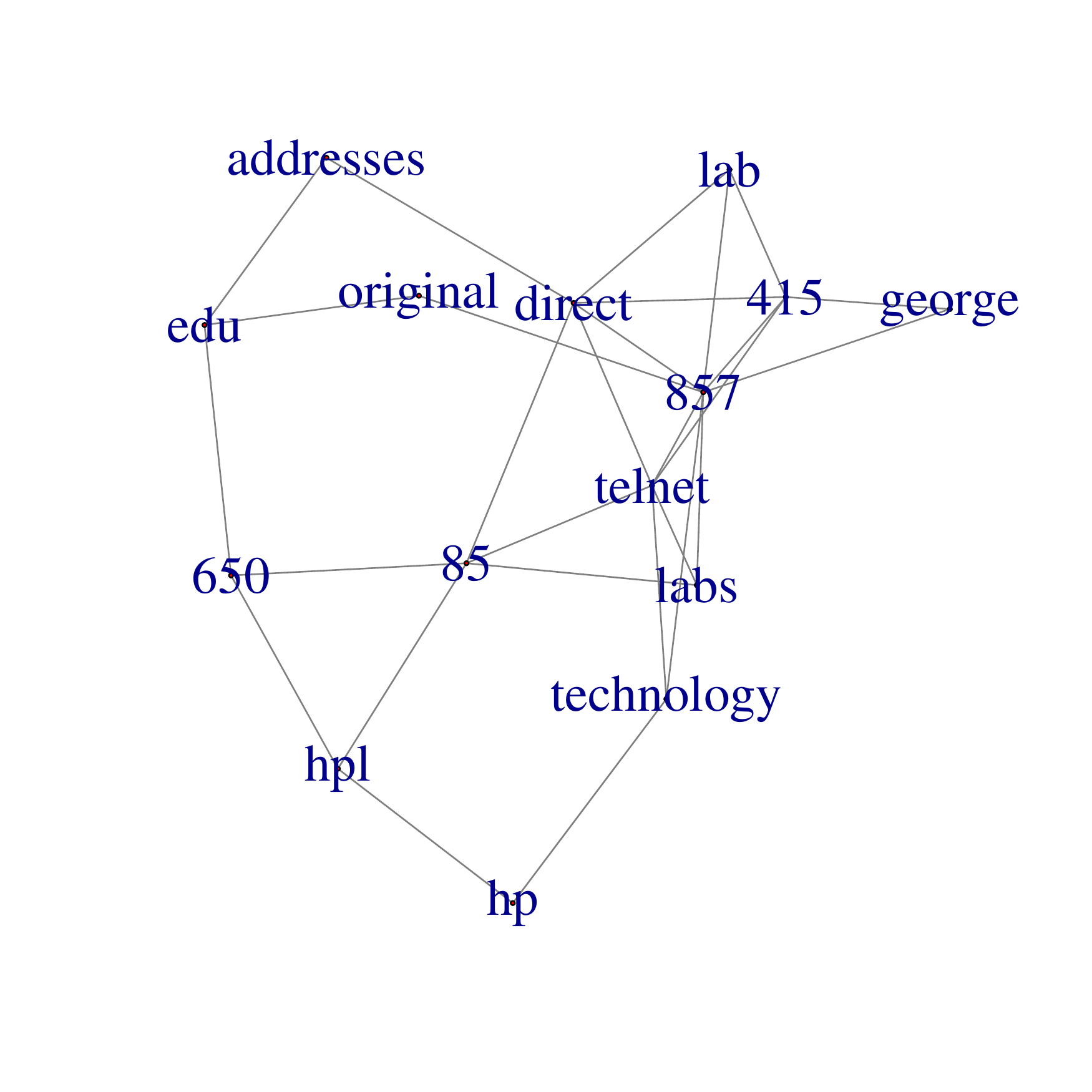,width=2.25in,height=2.5in,angle=0}
		\end{tabular}
	}
	\caption{The differential network for spam emails data set.}
	\label{fig2}
\end{figure}

We standardize the data and use a non-paranormal transformation to relax the assumption of Gaussian distribution. Figure (\ref{fig2}) illustrates the estimator given by our algorithm, where each node represents a specific feature. Our method indicates the existence of several hub features, including ``direct", ``telnet", ``technology", ``labs" and ``hp". Therefore, there might exist covariance structure changes between spam and non-span emails. For example, since the data is donated by Hewlett-Packard Labs, the words ``telnet", ``hp" and ``technology" will have a higher frequency in non-spam emails which means these features can help researchers to label the emails.

\subsection{Hepatocellular Carcinoma Data Set }

As a second example, we apply our algorithm to mRNA expression data of liver cancer patients from International Cancer Genome Consortium which is available at \url{https://icgc.org/icgc/cgp/66/420/824}. Several pathways from the KEGG pathway database (Ogata et al., 1999; Kanehisa et al., 2012) were studied
to determine the conditional dependency relationships  between liver cancers and normal patterns.  


To deal with the original data, we perform three steps. Firstly, we constrain the mRNAs in the following pathway: Pathways in cancer(05200), Transcriptional misregulation in cancer(05202), Viral carcinogenesis(05203), Chemical carcinogenesis(05204), Proteoglycans in cancer(05205), MicroRNAs in cancer(05206), Central carbon metabolism in cancer(05230), Choline metabolism in cancer(05231), and Hepatocellular carcinoma(05225). Secondly, we use the impute function from the R impute package to fill out the missing values. Thirdly, we standardize the data and use a non-paranormal transformation. This left us with 223 liver cancer patients and 222 normal patients with 1209 mRNAs in all.

\begin{figure}[!ht]
	\centerline{
		\begin{tabular}{cc}
			\psfig{figure=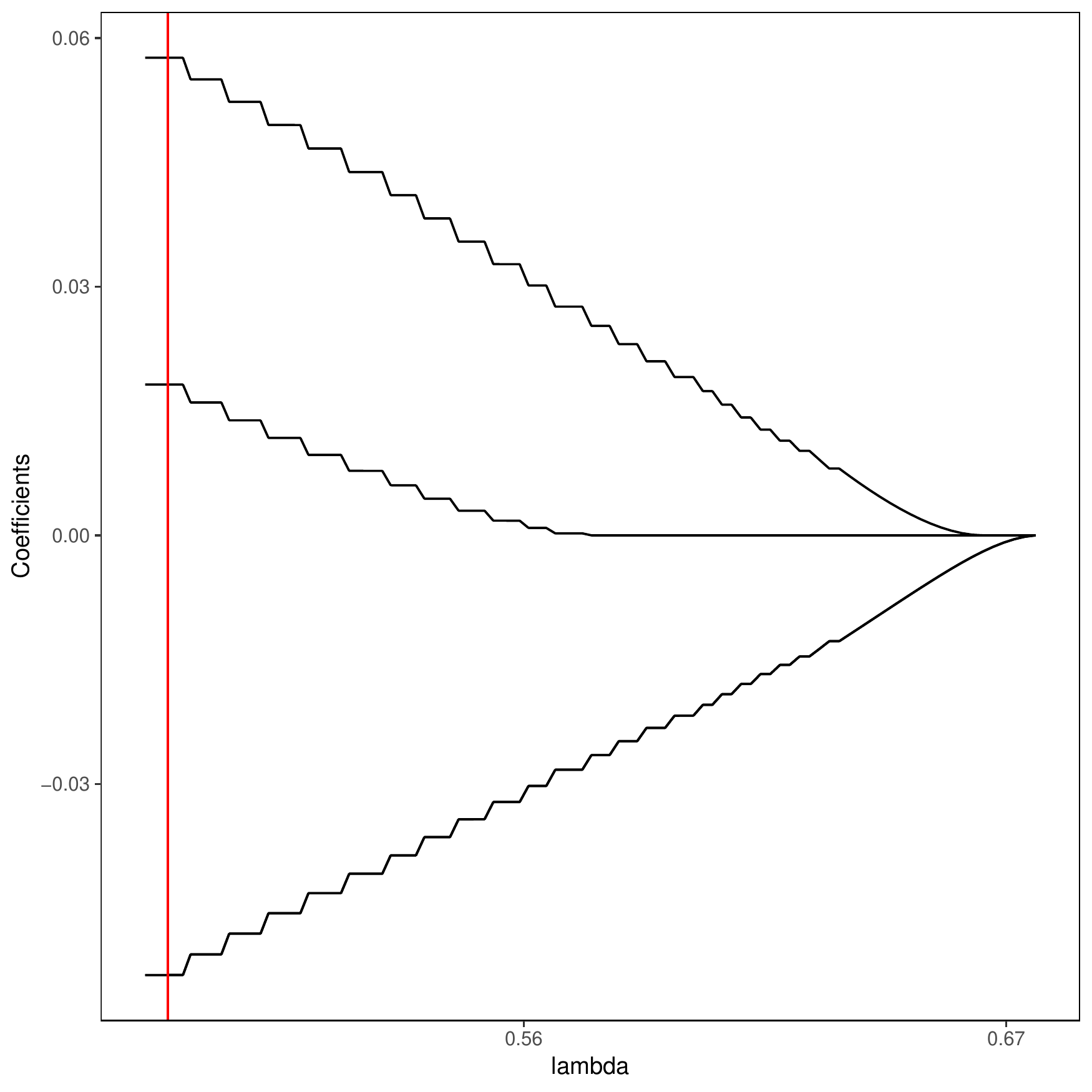,width=2.25in,height=2.5in,angle=0}&
			\psfig{figure=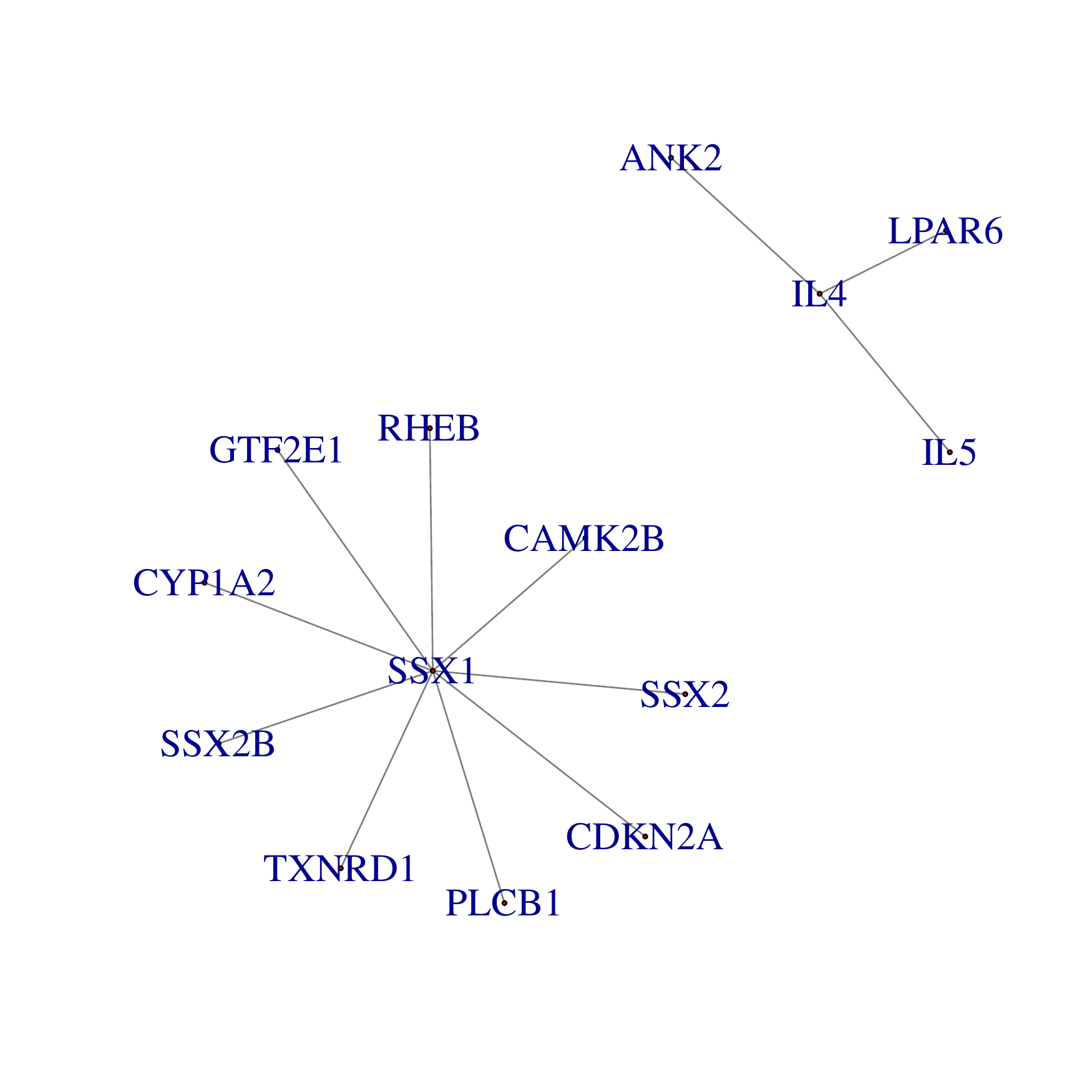,width=2.25in,height=2.5in,angle=0} 
		\end{tabular}
	}
	\caption{The differential network for Hepatocellular carcinoma data set .}
	\label{fig3}
\end{figure}

Figure (\ref{fig3}) summarizes the estimation given by our algorithm, where each node represents a specific mRNA. This figure show that real transcription networks often contain hub nodes. Our method indicates that SSX1 is an important mRNA. Indeed, SSX1 is a valid treatment option for CTNNB1 mutation positive HCC patients, while CTNNB1 is one of major mutations. Moreover, SSX1 as an oncogene is functionally validated \citep{Ding2014Transcriptomic}.

\section*{Acknowledgments}
Yu is supported in part by 2016YFC0902403 of Chinese Ministry of Science and Technology, and by National Natural Science Foundation of China 11671256.
Wang is partially supported by Shanghai Sailing Program 16YF1405700 and National Natural Science Foundation of China 11701367.

\section*{Appendix}
By the main results of \cite{beck2009fast}, to complete the proof of the Theorem 1, we only need to show that the loss function $L_1(\Delta)$ is convex which is the results of the following lemma.
\begin{lem}
	\label{lemma 1}
	The loss function (\ref{dloss}) is a smooth convex function, and its gradient is Lipschitz continuous with Lipschitz constant $L=\lambda_{\max}(S_1) \lambda_{\max}(S_2)$, that is
	\begin{align*}
	\| \nabla L_1(\Delta_1) - \nabla L_1(\Delta_2)\|_2 \leq L \| \Delta_1 - \Delta_2\|_2,
	\end{align*}
	where $\lambda_{\max}(S_i)$ is the largest eigenvalue of the sample covariance matrix $S_i$ for $i=1,2$. 
\end{lem}
\textbf{Proof}: Since the loss function (\ref{dloss}) is defined by 
\begin{align*}
L_1(\Delta)= \frac{1}{2} \tr\{\Delta \trans S_1 \Delta S_2\}- \tr\{\Delta (S_1-S_2)\},
\end{align*}
we can calculate the gradient of $L_1(\Delta)$ 
\begin{align*}
\nabla L_1(\Delta) = S_1\Delta S_2 - (S_1-S_2),
\end{align*}
and the Hessian matrix is $S_2 \otimes S_1$.
Since both covariance matrices $S_1$ and $S_2$ are definite positive matrix, the Hessian matrix is a definite positive matrix. Hence, the loss function $L_1(\Delta)$ is a smooth convex function.

Moreover, for any $\Delta_1, \Delta_2 \in \text{dom}(\nabla L_1)$, we have	
\begin{align*}
\| \nabla L_1(\Delta_1) - \nabla L_1(\Delta_2)\|_2&=\| S_1 (\Delta_1-\Delta_2) S_2 \|_2 \\
&=\| (S_2 \otimes S_1) \wvec(\Delta_1-\Delta_2) \|_2 \\
&\leq \lambda_{\max}(S_2 \otimes S_1) \|\wvec(\Delta_1-\Delta_2)\|_2 \\
&=\lambda_{\max}(S_1) \lambda_{\max}(S_2) \| \Delta_1-\Delta_2\|_2 .
\end{align*}
The proof is now completed.  \hfill$\fbox{}$
\bibliographystyle{abbrvnat}
\bibliography{ref}

\end{document}